\documentclass[aps,twocolumn]{revtex4}
\usepackage{amsfonts}

\input{tcilatex}

\begin{document}

\title{Compactons versus Solitons}
\author{Paulo E.G. Assis and Andreas Fring}

\begin{abstract}
We investigate whether the recently proposed $\mathcal{PT}$-symmetric
extensions of generalized Korteweg-de Vries equations admit genuine soliton
solutions besides compacton solitary waves. For models which admit stable
compactons having a width which is independent of their amplitude and those
which possess unstable compacton solutions the Painlev\'{e} test fails, such
that no soliton solutions can be found. The Painlev\'{e} test is passed for
models allowing for compacton solutions whose width is determined by their
amplitude. Consequently these models admit soliton solutions in addition to
compactons and are integrable.
\end{abstract}

\date{\today}
\maketitle

\preprint{HEP/123-qed}

\affiliation{Centre for Mathematical Science, City University London, Northampton Square,
London EC1V 0HB, UK}

\section{Introduction}

In a recent investigation Bender, Cooper, Khare, Mihaila and Saxena \cite
{comp1} have found compacton solutions, i.e. solitary wave solutions with
compact support, for $\mathcal{PT}$-symmetric extensions of generalized
Korteweg-de Vries (KdV) equations. The proposed models generalize various
systems previously studied and are described by the Hamiltonian density
\begin{equation}
\mathcal{H}_{l,m,p}=-\frac{u^{l}}{l(l-1)}-\frac{g}{m-1}u^{p}(iu_{x})^{m}.
\label{1}
\end{equation}
The density $\mathcal{H}_{l,2,p}$ reduces to a modification of a Hamiltonian
description \cite{comp2,comp3} of generalized KdV-equations \cite{comp4},
which are known to admit compacton solutions. For $l=3$, $p=0$ and $%
m=\varepsilon +1$ one obtains a re-scaled version of the $\mathcal{PT}$%
-symmetric extension of the KdV-equation ($\varepsilon =1$) introduced in 
\cite{AFKdV}. The first $\mathcal{PT}$-symmetric extensions of the
KdV-equation proposed in \cite{BBCF} can not be obtained from (\ref{1}) as
they correspond to non-Hamiltonian systems.

The virtue of $\mathcal{PT}$-symmetry, i.e. invariance under a simultaneous
parity transformation $\mathcal{P}:x\rightarrow -x$ and time reversal $%
\mathcal{T}:t\rightarrow -t,i\rightarrow -i$, for a classical Hamiltonian is
that it guarantees the reality of the energy due to its anti-linear nature 
\cite{AFKdV}. When quantizing $\mathcal{H}$ one also needs to ensure $%
\mathcal{PT}$-symmetry of the corresponding wavefunctions in order to obtain
real spectra \cite{EW,Bender:1998ke,SW,Bender:2003ve}. The most natural way
to implement $\mathcal{PT}$-symmetry in (\ref{1}) is to keep the
interpretation from the standard KdV-equation and view the field $u$ as a
velocity, such that it transforms as $u\rightarrow u$. Then $\mathcal{H}%
_{l,m,p}$ is $\mathcal{PT}$-symmetric for real coupling constant $g$ and all
possible real values of $l,m,p$. Alternatively, we could also allow a purely
complex coupling constant, i.e. $g\in i\Bbb{R}$, by transforming the field
as $u\rightarrow -u$, such that $\mathcal{H}_{l,m,p}$ is $\mathcal{PT}$%
-symmetric when $l$ is even and $p+m$ odd. For general reviews on $\mathcal{%
PT}$-symmetry and non-Hermitian Hamiltonian systems see \cite
{CArev,Benderrev,Alirev}.

The equation of motion resulting from the variational principle 
\begin{equation}
u_{t}=\left( \frac{\delta \int \mathcal{H}dx}{\delta u}\right)
_{x}=\sum\limits_{n=0}^{\infty }(-1)^{n}\left( \frac{d^{n}}{dx^{n}}\frac{%
\partial \mathcal{H}}{\partial u_{nx}}\right) _{x}  \label{var}
\end{equation}
for the Hamiltonian density $\mathcal{H}_{l,m,p}$ in (\ref{1}) is 
\begin{eqnarray}
&&u_{t}+u^{l-2}u_{x}+gi^{m}u^{p-2}u_{x}^{m-3}\left[ p(p-1)u_{x}^{4}\right.
\quad  \label{equm} \\
&&\left. +2pmuu_{x}^{2}u_{xx}+m(m-2)u^{2}u_{xx}^{2}+mu^{2}u_{x}u_{xxx}\right]
=0.  \nonumber
\end{eqnarray}

The main aim of this manuscript is to investigate whether this equation
admits soliton solutions and is therefore integrable for some specific
choices of the parameters $l,m,p$. We will also address the question of
whether it is possible to find solitons and compactons in the same model or
whether only one type of solutions may exist. To answer these questions one
could of course construct explicitly the soliton solutions, conserved
charges, Lax pairs, Dunkl operators, etc., which is usually a formidable
task. Instead we will carry out the Painlev\'{e} test following a proposal
originally made by Weiss, Tabor and Carnevale \cite{Pain1}. The test
provides an indication about the existence of soliton solutions and
discriminates between models, which are integrable those which are not.

\section{The Painlev\'{e} test}

The basic assumption for the existence of a soliton solution is that it
acquires the general form of a so-called Painlev\'{e} expansion \cite{Pain1} 
\begin{equation}
u(x,t)=\sum\limits_{k=0}^{\infty }\lambda _{k}(x,t)\phi (x,t)^{k+\alpha }.
\label{u}
\end{equation}
One further demands that in the limit $\phi (x,t)\rightarrow 0$, the
function $u(x,t)$ is meromorphic, such that the leading order singularity $%
\alpha $ is a negative integer and the $\ \lambda _{k}(x,t)$ are analytic
functions. The general procedure of the Painlev\'{e} test consists in
substituting the expansion (\ref{u}) into the equation of motion, (\ref{equm}%
) for the case at hand, and determining the functions $\lambda _{k}(x,t)$
recursively. A partial differential equation is said to pass the
Painlev\'{e} test when all $\lambda _{k}(x,t)$ can be computed, including
enough free parameters to match the order of the differential equation. In 
\cite{PainPA} we recently applied this method to $\mathcal{PT}$-symmetric
extensions of Burgers and the standard KdV-equation, where more details on
the generalities and literature may be found.

\subsection{Leading order singularities}

The Painlev\'{e} test stays and falls with the possibility that the initial
condition $\lambda _{0}$ can be determined, which is essential to commence
the iterative procedure to solve the recurrence relation. We compute $%
\lambda _{0}$ by substituting the first term in the expansion (\ref{u}),
i.e. $u(x,t)\rightarrow \lambda _{0}(x,t)\phi (x,t)^{\alpha }$, into (\ref
{equm}) and evaluating the values for all possible leading order
singularities $\alpha $. The individual terms in (\ref{equm}) have the
following leading order behaviour: $u_{t}\sim \phi ^{\alpha -1}$, $%
u^{l-2}u_{x}\sim \phi ^{\alpha (l-1)-1}$ and all remaining terms are
proportional to $\phi ^{\alpha (m+p-1)-m-1}$. Therefore the leading order
terms may only be cancelled if any of the following three conditions hold:

\begin{description}
\item[$i)$]  $\alpha -1=\alpha (l-1)-1\leq \alpha (m+p-1)-m-1,$

which results from assuming that $u_{t}$ and $u^{l-2}u_{x}$ constitute the
leading order terms. In this case we obtain $l=2$ and the inequality $\alpha
(2-m-p)\leq -m$. Thus $\alpha $ remains undetermined.

\item[$ii)$]  $\alpha -1=\alpha (m+p-1)-m-1\leq \alpha (l-1)-1,$

which corresponds to the assumption that $u^{l-2}u_{x}$ is the least
singular term and matching the leading orders of all the remaining ones.
Then we conclude that $l\leq 2$ and $\alpha $ is fixed to $\alpha =m/(m+p-2)$%
.

\item[$iii)$]  $\alpha (l-1)-1=\alpha (m+p-1)-m-1\leq \alpha -1.$

which is the consequence of $u_{t}$ being least singular term and the
matching of the remaining ones. This means the leading order singularity of $%
u(x,t)$ is of the order 
\begin{equation}
\alpha =\frac{m}{p+m-l}\in \Bbb{Z}^{-}\quad \text{and \quad }l\geq 2.
\label{alpha}
\end{equation}
Cancelling the leading order terms then yields
\begin{equation}
\lambda _{0}^{(n)}=e^{2\pi in\alpha /m}[gl(l-1)]^{-\alpha /m}(i\alpha \phi
_{x})^{-\alpha },
\end{equation}
where $1\leq n\leq p+m-l$ indicates the different roots of the determining
equation.
\end{description}

In principle we could also envisage a scenario in which $u_{t}$ and $%
u^{l-2}u_{x}$ are the least dominant terms and the leading order singularity
is cancelled by all the remaining terms. However, all these terms only
differ by an overall numerical factor, such that $\lambda _{0}$ turns out to
be zero in this case and we can therefore discard this case.

\subsection{Resonances}

A key feature of the Painlev\'{e} test is the occurrence of so-called
resonances, which arise whenever the coefficient in front of a specific $%
\lambda _{r}$ in the recurrence relations becomes zero. This implies that $%
\lambda _{r}$ can not be determined recursively. When in this case the
remaining part of the recurrence relation becomes an identity, the $\lambda
_{r}$ becomes a free parameter, otherwise the Painlev\'{e} test fails. The
possible values for $r$ can be found by substituting 
\begin{equation}
u(x,t)\rightarrow \lambda _{0}(x,t)\phi (x,t)^{\alpha }+\lambda _{r}\phi
(x,t)^{r+\alpha }  \label{vv}
\end{equation}
into (\ref{equm}) and computing all possible values of $r$ for which $%
\lambda _{r}$ becomes a free parameter. Considering the case $iii)$ for
integer values $l,m,p$ the coefficients of the leading order $\phi
^{r+\alpha (l-1)-1}$ is proportional to
\begin{equation}
\lambda _{r}g^{\alpha (2-l)/m}(r+1)(r+\alpha l)[r+\alpha (l-1)]\phi
_{x}^{\alpha (2-l)+1}.  \label{res}
\end{equation}
This means that besides the so-called fundamental resonance at $r=1$, we
also find two more resonances at $r=-\alpha l,\alpha (1-l)$. Since the
differential equation (\ref{equm}) is of order three all these models fully
pass the Painlev\'{e} test provided $\lambda _{-\alpha l}$ and $\lambda
_{\alpha (1-l)}$ can indeed be chosen freely.

The standard procedure to verify this would be now to derive the recursive
equation resulting from combining (\ref{u}) and (\ref{equm}). Clearly for
generic values of $l,m,p$ this will be extremely lengthy, but even for
specific choices it is fairly complicated. It suffices, however, to compute
the $\lambda _{k}$ up to $k>-\alpha l$. We will present these values for
various examples for several choices of the parameters $l,m,p$ corresponding
to scenarios leading to solutions with qualitatively different kinds of
behaviour.

\section{Generalized KdV-equation}

Cooper, Khare and Saxena \cite{comp5} found that in the generalized KdV
equation, i.e. $m=2$, a necessary condition for compactons to be stable is
to consider models with $2<l<p+6$. This means none of the conditions $i)$ or 
$ii)$ for the leading order singularity to cancel can be satisfied. The
special choice $l=p+2,0<p\leq 2$ guarantees that the compacton solutions
have in addition a width which is independent of their amplitude \cite{comp2}%
. For that particular case also the condition $iii)$ admits no solution,
such that the Painlev\'{e} test fails.

However, for models which admit stable compacton solutions having a width
depending on the amplitude we can find solutions to the condition $iii)$ and
proceed with the Painlev\'{e} test. For instance, $m=2,p=1,l=5$ is such a
choice. In this case we find from (\ref{alpha}) that $\alpha =-1$ and the
leading order singularity of the corresponding differential equation is $%
\phi ^{-5}$. Computing now order by order the functions $\lambda _{k}$ we
find the two solutions 
\begin{eqnarray}
\lambda _{0}^{\pm } &=&\pm 2i\sqrt{5g}\phi _{x}\text{, \quad }\lambda
_{1}^{\pm }=\mp i\sqrt{5g}\frac{\phi _{xx}}{\phi _{x}}\text{, }  \nonumber \\
\lambda _{2}^{\pm } &=&\mp i\frac{\sqrt{5g}}{6}\frac{\left( 3\phi
_{xx}^{2}-2\phi _{x}\phi _{xxx}\right) }{\phi _{x}^{3}}\text{,} \\
\lambda _{3}^{\pm } &=&\frac{3\phi _{t}\phi _{x}^{2}\mp 4i\sqrt{5g^{3}}%
\left( 6\phi _{xx}^{3}-6\phi _{x}\phi _{3x}\phi _{xx}+\phi _{x}^{2}\phi
_{4x}\right) }{48g\phi _{x}^{5}}.  \nonumber
\end{eqnarray}
Crucially we observe next that $\lambda _{4}^{\pm }$ and $\lambda _{5}^{\pm
} $ can be chosen arbitrarily. The remaining $\lambda _{k}^{\pm }$ for $k>5$
can all be computed, but the expressions are all extremely cumbersome and we
will therefore not report them here. Making, however, the further assumption
on $\phi $ to be a travelling wave, i.e. $\phi (x,t)=x-\omega t$, simplifies
the expressions considerably. Choosing $\lambda _{4}^{\pm }=\lambda
_{5}^{\pm }=0$ the two solutions for that scenario reduce to
\begin{eqnarray}
\lambda _{3\kappa +1}^{\pm } &=&\lambda _{3\kappa +2}^{\pm }=0\quad \quad 
\text{for }\kappa =0,1,2,\ldots  \nonumber \\
\lambda _{0}^{\pm } &=&\pm 2i\sqrt{5g}\text{, }\lambda _{3}^{\pm }=-\frac{%
\omega }{16g}\text{, }  \nonumber \\
\lambda _{6}^{\pm } &=&\mp \frac{3i\omega ^{2}}{3584\sqrt{5}g^{5/2}}\text{, }%
\lambda _{9}^{\pm }=\frac{\omega ^{3}}{573440g^{4}}\text{, } \\
\lambda _{12}^{\pm } &=&\pm \frac{33i\omega ^{4}}{1669857280\sqrt{5}g^{11/2}}%
\text{, }  \nonumber \\
\lambda _{15}^{\pm } &=&-\frac{3\omega ^{5}}{66794291200g^{7}}\text{, \ldots 
}  \nonumber
\end{eqnarray}
We conclude that the Painlev\'{e} test is passed for this choices of
parameters, which means that besides stable compacton solutions, whose width
depends on their amplitude, we also find genuine solitons in these models
and, provided the series (\ref{u}) converges, they are therefore integrable.

In the unstable compacton regime, i.e. $l\leq 2$ or $l\geq p+6$, the
condition $iii)$ can not be satisfied. Consequently we do not expect to find
genuine soliton solutions. We have also verified this type of behaviour for
other representative examples which we do not present here.

\section{$\mathcal{PT}$-symmetric generalized KdV-equation}

For the $\mathcal{PT}$-symmetric extensions of the generalized KdV-equation (%
\ref{equm}) the necessary condition for compactons to be stable was extended
by Bender et al \cite{comp1} to $2<l<p+3m$. Thus also for generic values of $%
m$ none of the conditions $i)$ or $ii)$ for the leading order singularity to
cancel can be satisfied. Furthermore, the requirement for stable compacton
solutions to possess also a width which is independent of their amplitude
was generalized in \cite{comp1} to $l=p+m$. As for the special case $m=2$
this value coincides with the leading order singularity resulting from the
condition $iii)$ tending to infinity and therefore the Painlev\'{e} test
fails.

As in the previous case, for models which have stable compacton solutions
whose width is a function of their amplitude the Painlev\'{e} test has a
chance to pass, as one can find a value for the leading order singularity
and potentially has the correct amount of resonances. We verify this for the
example $m=3,p=1,l=7$, for which we obtain $\alpha =-1$ and $\phi ^{-7}$ as
the leading order singularity in (\ref{equm}). Since $-\alpha /m=1/3$ in
this case, we find now three non-equivalent solutions related to the
different roots for the $\lambda _{k}^{(n)}$ with $n=1,2,3$, of which the
first terms are 
\begin{eqnarray}
\lambda _{0}^{(n)} &=&-ie^{2\pi in/3}(42g)^{1/3}\phi _{x}\text{, }  \nonumber
\\
\lambda _{1}^{(n)} &=&\frac{ie^{2\pi in/3}(21g)^{1/3}\phi _{xx}}{2^{2/3}\phi
_{x}}\text{, } \\
\lambda _{2}^{(n)} &=&\frac{ie^{2\pi in/3}(7g)^{1/3}\left( 3\phi
_{xx}^{2}-2\phi _{x}\phi _{xxx}\right) }{2(6)^{2/3}\phi _{x}^{3}}\text{,} 
\nonumber \\
\lambda _{3}^{(n)} &=&\frac{ie^{2\pi in/3}(7g)^{1/3}\left( 6\phi
_{xx}^{3}-6\phi _{x}\phi _{xxx}\phi _{xx}+\phi _{x}^{2}\phi _{xxxx}\right) }{%
4(6)^{2/3}\phi _{x}^{5}}.  \nonumber
\end{eqnarray}
From (\ref{res}) we know that we should encounter resonances at the level $6$
and $7$, which is indeed the case as we find that $\lambda _{6}^{(n)}$and $%
\lambda _{7}^{(n)}$ can be chosen freely. The remaining $\lambda _{k}^{(n)}$
for $k>7$ can all be computed iteratively and the Painlev\'{e} test is
passed for this example.

For a travelling wave ansatz $\phi (x,t)=x-\omega t$ with the choice $%
\lambda _{6}^{(n)}=$ $\lambda _{7}^{(n)}=0$ the expressions simplify to 
\begin{eqnarray}
\lambda _{5\kappa +1}^{(n)} &=&\lambda _{5\kappa +2}^{(n)}=\lambda _{5\kappa
+3}^{(n)}=\lambda _{5\kappa +4}^{(n)}=0,\text{~for }\kappa =0,1,\ldots 
\nonumber \\
\lambda _{0}^{(n)} &=&-ie^{2\pi in/3}(42g)^{1/3}\text{, }\lambda _{5}^{(n)}=%
\frac{e^{4\pi in/3}\omega }{36(42)^{1/3}g^{4/3}}\text{,} \\
\lambda _{10}^{(n)} &=&\frac{17i\omega ^{2}}{598752g^{3}}\text{, }  \nonumber
\\
\lambda _{15}^{(n)} &=&-\frac{53e^{\frac{2in\pi }{3}}\omega ^{3}}{%
21555072(42)^{2/3}g^{14/3}}\text{,\ldots }  \nonumber
\end{eqnarray}
Thus we observe no qualitative difference in the $\mathcal{PT}$-symmetric
extensions in comparison to the case $m=2$ and find that also in this one
may have stable compacton solutions, whose width depends on their amplitude
and genuine solitons at the same time.

In the unstable compacton regime, that is $l\leq 2$ or $l\geq p+3m$, the
condition $iii)$ can not be satisfied and the Painlev\'{e} test fails. Once
again we do not represent here other representative examples for which we
obtained the same type of behaviour.

\section{Deformations of Burgers equation}

Considering $m=1,p=1,l=3$ in the equation of motion (\ref{equm}) is a very
simple example leading to a Painlev\'{e} expansion for $u(x,t)$, which can
even be truncated after the second term. As this type of behaviour is
reminiscent of B\"{a}cklund transformation generating solutions found in
other models \cite{Pain1}, we present this case briefly. For this choice (%
\ref{equm}) simply reduces to
\begin{equation}
u_{t}+uu_{x}-2igu_{xx}+\frac{igu}{u_{x}^{2}}\left(
u_{xx}^{2}-u_{x}u_{xxx}\right) =0,  \label{111}
\end{equation}
which can be viewed as a deformation of Burgers equation \cite{PainPA}
corresponding to the first three terms. Proceeding as in the previous
sections, we find the solution
\begin{equation}
u(x,t)=\frac{-6ig\phi _{x}}{\phi }+\frac{6ig\phi _{xx}-3\phi _{t}}{2\phi _{x}%
}
\end{equation}
provided that $\phi $ satisfies the equation 
\begin{equation}
\phi _{x}^{2}\phi _{tt}+\phi _{t}^{2}\phi _{xx}=2\phi _{tx}\phi _{t}\phi
_{x}.  \label{xc}
\end{equation}
A travelling wave $\phi (x,t)=x-\omega t$ is for instance a solution of (\ref
{xc}), such that we obtain the simple expression
\begin{equation}
u(x,t)=\frac{6ig\phi _{x}}{\omega t-x}+\frac{3}{2}\omega   \label{so}
\end{equation}
for the solution of (\ref{111}). Incidentally, the travelling wave solution
for Burger's equation \cite{Pain1} coincides with (\ref{so}).

\section{Conclusions}

In previous investigations \cite{comp2,comp3,comp5} various criteria have
been found, which separate the models $\mathcal{H}_{l,m,p}$ into three
distinct classes exhibiting qualitatively different types of compacton
solutions, unstable compactons and stable compactons, which have either
dependent or freely selectable width $A$ and amplitude $\beta $. We have
carried out the Painlev\'{e} test for various examples for each of these
classes and found that all models which allow stable compactons for which
the width can not be chosen independently from their amplitude pass the
Painlev\'{e} test. Assuming that the Painlev\'{e} expansion (\ref{u})
converges these models possess the Painlev\'{e} property \cite{Gramma} and
allow therefore for genuine soliton solutions and are thus integrable. We
found that the generalized KdV equation resulting from $\mathcal{H}_{l,2,p}$
and their $\mathcal{PT}$-symmetric extensions $\mathcal{H}_{l,m,p}$ have the
same qualitative behaviour in the three different regimes. For convenience
we summarize the different qualitative behaviours in the following table:

\begin{center}
\begin{tabular}{|l||l|l|}
\hline
$\mathcal{H}_{l,m,p}$ & compactons & solitons \\ \hline
$l=p+m$ & stable, dependent $A,\beta $ & no \\ \hline
$2<l<p+3m$ & stable, independent $A,\beta $ & yes \\ \hline
$l\leq 2$ or $l\geq p+3m$ & unstable & no \\ \hline
\end{tabular}
\end{center}

\noindent Table 1: The models $\mathcal{H}_{l,m,p}$ and their solutions.

Clearly our investigations do not constitute a full fletched mathematical
proof as we based our findings on various representative examples for the
different classes and it would be very interesting to settle this issue more
rigorously with a generic argumentation not relying on case-by-case studies.
At the same time such a treatment would probably provide a deeper
understanding about the separation of the different models. Nonetheless, our
findings provide enough evidence to make it worthwhile to investigate the
models which pass the test with other techniques developed in the field of
integrable models, whereas models which do not pass the test may be excluded
from such investigations.

\medskip \noindent \textbf{Acknowledgments}: P.E.G.A. is supported by a City
University London research studentship.


\end{document}